\documentclass[aps,prx,reprint,floatfix,groupedaddress,showpacs,superscriptaddress,amsmath,amssymb]{revtex4-1}
\usepackage{graphicx}
\usepackage{units}
\usepackage{amsmath}
\usepackage{amsfonts}
\usepackage{mathrsfs}
\usepackage{amsbsy}
\usepackage{bm}
\usepackage[hypertexnames=false,linktocpage=true,colorlinks=true,linkcolor=blue,anchorcolor=blue,citecolor=blue,filecolor=blue,urlcolor=blue,bookmarksnumbered=true,pdfview=FitB,breaklinks=true]{hyperref}

\newcommand{\hh}{\ensuremath{H}}
\newcommand{\kk}{\ensuremath{K}}

\begin{document}

\title{Structural and antiferromagnetic properties of Ba(Fe$_{1-x-y}$Co$_x$Rh$_y$)$_2$As$_2$ compounds}

\author{M.~G.~Kim}\email{mgkim@lbl.gov}
\affiliation{Materials Sciences Division, Lawrence Berkeley National Laboratory, Berkeley, California 94720, USA}
\author{T.~W.~Heitmann}
\affiliation{The Missouri Research Reactor, University of Missouri, Columbia, Missouri 65211, USA}
\author{S. R. Mulcahy}\thanks{ Current Affiliation/Address: Geology Department, Western Washington University, Bellingham, WA, 98225, USA.  (email sean.mulcahy@wwu.edu)}
\affiliation{\mbox{Department~of~Earth~and~Planetary~Science,~University~of~California,~Berkeley,~California~94720,~USA}}
\author{E.~D.~Bourret-Courchesne}
\affiliation{Materials Sciences Division, Lawrence Berkeley National Laboratory, Berkeley, California 94720, USA}
\author{R.~J.~Birgeneau}
\affiliation{Materials Sciences Division, Lawrence Berkeley National Laboratory, Berkeley, California 94720, USA}
\affiliation{\mbox{Department~of~Materials~Science~and~Engineering,~University~of~California,~Berkeley,~California~94720,~USA}}
\affiliation{Department of Physics, University of California, Berkeley, California 94720, USA}

\date{\today}

\begin{abstract}
We present a systematic investigation of the electrical, structural, and antiferromagnetic properties for the series of Ba(Fe$_{1-x-y}$Co$_{x}$Rh$_{y}$)$_{2}$As$_{2}$ compounds with fixed $x \approx$ 0.027 and $ 0 \leq y \leq 0.035$. We compare our results for the Co-Rh doped Ba(Fe$_{1-x-y}$Co$_{x}$Rh$_{y}$)$_{2}$As$_{2}$ compounds with the Co doped Ba(Fe$_{1-x}$Co$_{x}$)$_{2}$As$_{2}$ compounds. We demonstrate that the electrical, structural, antiferromangetic, and superconducting properties of the Co-Rh doped compounds are similar to the properties of the Co doped compounds. We find that the overall behaviors of Ba(Fe$_{1-x-y}$Co$_{x}$Rh$_{y}$)$_{2}$As$_{2}$ and Ba(Fe$_{1-x}$Co$_{x}$)$_{2}$As$_{2}$ compounds are very similar when the total number of extra electrons per Fe/$TM$ ($TM$ = transition metal) site is considered, which is consistent with the rigid band model. Despite the similarity, we find that the details of the transitions, for example, the temperature difference between the structural and antiferromagnetic transition temperatures and the incommensurability of the antiferromangetic peaks, are different between Ba(Fe$_{1-x-y}$Co$_{x}$Rh$_{y}$)$_{2}$As$_{2}$ and Ba(Fe$_{1-x}$Co$_{x}$)$_{2}$As$_{2}$ compounds.
\end{abstract}


\maketitle

\section{Introduction}

The high-temperature superconductivity in the FeAs-based compounds is closely related to the underlying structural and magnetic properties. 
The parent BaFe$_2$As$_2$ compound exhibits structural and antiferromagnetic (AFM) phase transitions.\cite{Huang08, Kim11, canfield10, Johnston10, Dai15} The structure changes from a tetragonal ($I4/mmm$) to an orthorhombic ($Fmmm$) structure.\cite{Huang08, Kim11, canfield10, Johnston10, Dai15} The AFM transition occurs at a temperature ($T_\mathrm{N}$) slightly lower than the structural transition temperature ($T_\mathrm{S}$) and the AFM ordering is commensurate and characterized by the propagation vector $\bm{Q}_\mathrm{AFM} = (1,~0,~1)$ in the orthorhombic notation.\cite{Huang08, Kim11, canfield10, Johnston10, Dai15} 

Superconductivity in this system can be effectively achieved by tuning external parameters.\cite{canfield10, Johnston10, Dai15} One of the parameters is doping by substituting transition metal elements for Fe.\cite{canfield10, Johnston10, Dai15,sefat08, ni08, Canfield09,  li_superconductivity_2009, ni_phase_2009, han_2009, saha_superconductivity_2010} This is noted as electron or hole doping since these elements are considered to possess additional carriers when compared with Fe. With electron doping by transition-metal elements, particularly in Ba(Fe$_{1-x}TM_x$)$_2$As$_2$ with $TM =$ Co\cite{sefat08, ni08}, Ni\cite{li_superconductivity_2009, Canfield09}, Rh\cite{ni_phase_2009, han_2009}, Pd\cite{ni_phase_2009, han_2009}, Ir\cite{han_2009}, or Pt\cite{saha_superconductivity_2010}, the structural and AFM transitions are continuously suppressed to lower temperatures and the difference between $T_\mathrm{S}$ and $T_\mathrm{N}$ becomes larger with increasing substitution levels. Superconductivity emerges at a sufficient doping level, usually before the complete suppression of those transitions.\cite{canfield10, Johnston10, Dai15,sefat08, ni08, li_superconductivity_2009, Canfield09, ni_phase_2009, han_2009, saha_superconductivity_2010}

With the emergence of superconductivity, the superconducting and antiferromagnetic states compete for the same quasiparticles. As a result, when superconductivity becomes dominant, the AFM ordering is weakened, which is observed as the suppression of the AFM order parameter below the superconducting transition temperature ($T_\mathrm{c}$).\cite{Kreyssig10, wang_electron-doping_2010, pratt_coexistence_2009, christianson_2009, Fernandes10} Since the crystal structure is coupled to the magnetism via the nematic order parameter\cite{Kim11, nandi_anomalous_2010}, the structure of the system also alters below $T_\mathrm{c}$. The orthorhombic structure becomes less orthorhombic below $T_\mathrm{c}$ and eventually re-enters to a tetragonal phase at higher doping levels.\cite{nandi_anomalous_2010}  

Detailed measurements of the AFM ordering by neutron diffraction also revealed that the commensurate (C) AFM order\cite{Kim_com_2010} becomes incommensurate (IC), $\bm{Q}_\mathrm{AFM} + \bm{\tau}$ with a small incommensurability $\bm{\tau}$, at higher substitution levels in Ba(Fe$_{1-x}TM_x$)$_2$As$_2$ with $TM =$ Co\cite{Pratt_2011} and Ni\cite{Kim12}. Because the C and IC AFM phases coexist in certain doping levels, the C-to-IC transition is first order.\cite{Pratt_2011, Kim12} In contrast, non-superconducting electron-doped Ba(Fe$_{1-x}$Cu$_x$)$_2$As$_2$ compounds do not show the C-to-IC transition while the suppression of the AFM ordering is similar to that in superconducting compounds.\cite{Kim12} Thus, not only the suppression of the AFM ordering but also the C-to-IC transition may be linked to the superconductivity in this system. 

Interestingly, a simple rigid band model can explain the properties of the electron-doped superconducting compounds, Ba(Fe$_{1-x}TM_x$)$_2$As$_2$.\cite{Canfield09, ni_phase_2009, Kim12} In the rigid band picture, Co gives one electron more than Fe and Ni gives two electrons more than Fe; Ni doping affects the properties of the compound twice as effectively as Co doping.\cite{Canfield09, ni_phase_2009, Kim12} When the phase diagrams of Ba(Fe$_{1-x}$Co$_x$)$_2$As$_2$ and Ba(Fe$_{1-x}$Ni$_x$)$_2$As$_2$ are plotted in terms of the number of extra electrons per the Fe/$TM$ site, those phase diagrams lie on top of each other.\cite{Canfield09, ni_phase_2009} Similarly, the rigid band picture is also valid for Rh, Pd, and other electron doping elements that induce superconductivity. However, previous studies show that electron or hole doping in Ba(Fe$_{1-x}TM_x$)$_2$As$_2$ with $TM =$ Cr\cite{marty_cr_2011}, Mn\cite{kim_2011, Thaler_mn_11}, or Cu\cite{Canfield09,Kim12} show different magnetic properties and no superconductivity; this behavior deviates from the rigid band prediction.  

Since the structural, magnetic, and superconducting properties are similar in compounds in which the rigid band approximation is valid, one can imagine that the doping effect plays a major role in the properties of the FeAs-based compounds while it has been argued that the aspects of the crystal structure, such as the pnictogen-Fe-pnictogen angle or the pnictogen height, may directly affect superconducting properties.\cite{Johnston10, Kuroki09, Calderon09} For example, although Co and Rh have different atomic/ionic sizes and thus are expected to change the crystal structure differently, the superconducting properties as well as the magnetic properties are quite similar in Ba(Fe$_{1-x}$Co$_x$)$_2$As$_2$ (``Co122") and Ba(Fe$_{1-x}$Rh$_x$)$_2$As$_2$ (``Rh122"),\cite{ni_phase_2009} which may be explained by a dominant electron doping effect. Hence, we can expect that simultaneous doping of Co and Rh in Ba(Fe$_{1-x-y}$Co$_x$Rh$_y$)$_2$As$_2$ (``CoRh122") compounds should exhibit similar magnetic and superconducting properties even if the details of its structure are different from those in Co122 or Rh122. 

Here, we present a systematic study of the electrical properties, lattice parameters, and structural and antiferromagnetic properties as a function of total doping level, $x + y$ in Ba(Fe$_{1-x-y}$Co$_x$Rh$_y$)$_2$As$_2$. We find that the details of the crystal structures, observed by the lattice parameters $a$ and $c$, are different in Co122, Rh122, and CoRh122 but the superconducting transition temperatures are similar in all these compounds. We show that the structural/AFM transitions, the AFM ordering, and their phase diagrams are quite similar in Ba(Fe$_{1-x-y}$Co$_x$Rh$_y$)$_2$As$_2$, Co122, and Rh122, indicating that the rigid band picture is applicable in explaining the properties of Co and Rh doubly doped Ba(Fe$_{1-x-y}$Co$_x$Rh$_y$)$_2$As$_2$ compounds.

\section{Experiment}

Single crystals of  Ba(Fe$_{1-x-y}$Co$_{x}$Rh$_{y}$)$_{2}$As$_{2}$ with $x \approx$ 0.027 were grown out of a Fe/Co/Rh-As flux using conventional high-temperature flux growth.  We fixed the content of the Co doping to be $x \approx 0.027$ and varied the Rh content, $y$. 
First, we prepared Fe/Co/Rh-As precursors with a ratio of Fe:Co:Rh:As = ($1-x-y$) : $x$ : $y$ : 1, which were sealed in an evacuated quartz tube. The prepared precursor powders were heated following the temperature steps described in Ref.~\onlinecite{Chen11}. 
Then the resulting precursor was mixed with Ba pieces in the ratio of Ba:Fe$_{1-x-y}$Co$_x$Rh$_y$As = 1 : 4, which was also sealed in an evacuated quartz tube. To grow single crystals, we applied the heating procedure described in Ref.~\onlinecite{ni08} and we used the centrifugal decanting method at 1000 $^\circ$C to separate crystals from the flux. 

\begin{figure}[t!]
\centering
\includegraphics[width=0.95\linewidth]{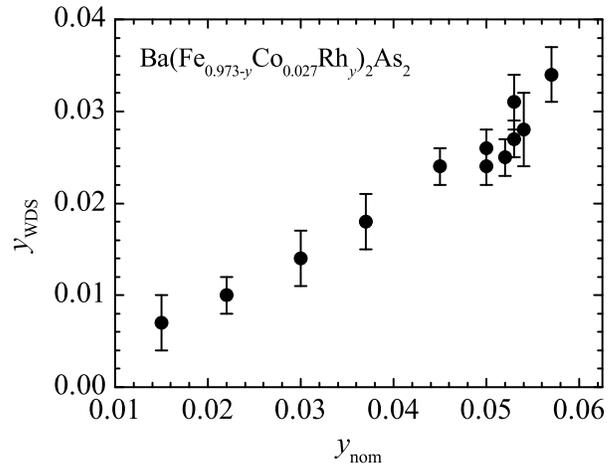} 
\caption{Measured Rh concentration vs nominal Rh concentration for the Ba(Fe$_{1-x-y}$Co$_x$Rh$_y$)$_2$As$_2$ with $x \approx 0.027$ compounds.}
\label{wds}
\end{figure}

Compositional analyses were acquired using a Cameca SX-51 electron microprobe equipped with 5 tunable wavelength dispersive spectrometers (WDS). Analyses were conducted with a 20 keV accelerating voltage, 20 nA beam current, and 5 micron beam diameter. Peak and background count times for all elements were 10 seconds. Analyses yielded a relative uncertainty less than 5\%. Figure~\ref{wds} shows a summary of the nominal Rh concentration vs. actual Rh concentration from the WDS measurements. We fix the nominal Co concentration, $x_\mathrm{nom} = 0.032$ which results in actual Co concentrations of $x_\mathrm{WDS} = 0.026-0.029$; this demonstrates that $y_\mathrm{WDS}$ increases roughly linearly with the nominal doping concentration for Ba(Fe$_{1-x-y}$Co$_x$Rh$_y$)$_2$As$_2$ with $y_\mathrm{WDS} \approx \frac{1}{2} y_\mathrm{nom}$.

Electrical transport data were collected by a Quantum Design Physical Property Measurement System 􏰅(PPMS)􏰆. Electrical contacts were made to the sample using Leitsilber 200 conductive silver paint to attach Au wires in a four-probe configuration. The measurements were done between $T = 2$ K and 300 K. 
Powder x-ray diffraction data were collected at room temperature with a Siemens diffractometer using Cu-K$\alpha_1$ radiation. Several small crystals from the same growth batch were collected and ground into powder for the measurements. The lattice parameters were obtained by the Le Bail extraction method using the Rietica program\cite{Rietica}.

For neutron diffraction measurements, single pieces of crystals with a typical mass of approximately 200 mg were selected from each growth batch. We performed the diffraction measurements at the TRIAX triple-axis spectrometer at the University of Missouri Research Reactor. The beam collimators before the monochromator, between the monochromator and sample, between the sample and analyzer, and between the analyzer and detector were $60'-40'-~$sample$~-40'-80'$ collimation. We used fixed $E_\mathrm{i} = E_\mathrm{f} = 14.7$ meV and two pyrolytic graphite filters, one before the analyzer and one before the monochromator to eliminate higher harmonics in the incident beam. 
Measurements were performed in a closed-cycle refrigerator between room temperature and the base temperature, $T \approx 5-7$ K of the refrigerator. We define $\bm{Q} = \left(\hh,\kk,L\right) = \frac{2\pi}{a}\hh\hat{\imath} + \frac{2\pi}{b}\kk\hat{\jmath} + \frac{2\pi}{c}L\hat{k}$ where the orthorhombic lattice constants are $a \geq b \approx 5.6 $\,\AA~and $c\approx 13 $\,\AA. 
Samples were studied in the vicinity of $\bm{Q}_\mathrm{AFM} = (1,~0,~3)$ in the  ($\zeta,~K,~3\zeta$) plane, allowing a search for incommensurability along the $\bm{b}$ axis ([$0,~K,~0$], transverse direction) as found for Ba(Fe$_{1-x}$Co$_x$)$_2$As$_2$\cite{Pratt_2011} and Ba(Fe$_{1-x}$Ni$_x$)$_2$As$_2$\cite{Kim12}. 
All samples exhibited small mosaicities, $\leq 0.4^{\circ}$ full-width-at-half-maximum (FWHM) measured by rocking scans, demonstrating high sample quality. 

\section{Results and discussion}

\begin{figure}[t!]
\centering
\includegraphics[width=0.95\linewidth]{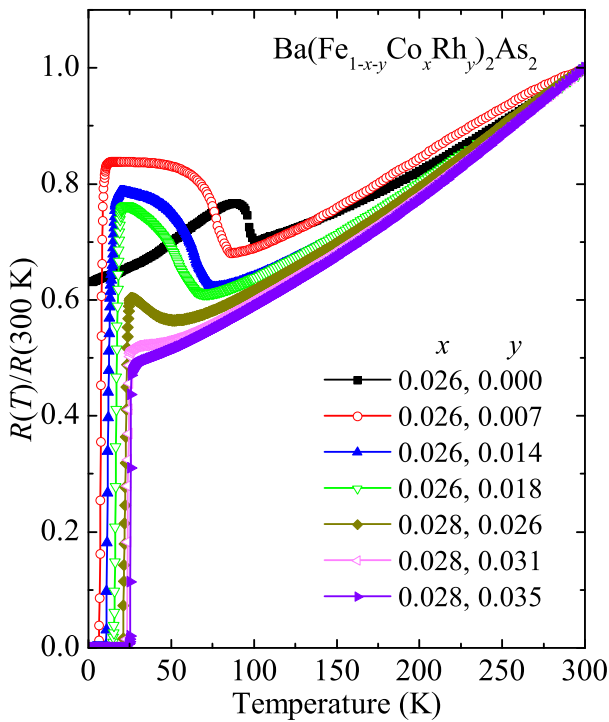} 
\caption{(color online) The temperature-dependent resistance, normalized by the room temperature value, for Ba(Fe$_{1-x-y}$Co$_x$Rh$_y$)$_2$As$_2$.}
\label{resistivity}
\end{figure}

\begin{figure}[t!]
\includegraphics[width=1.0\linewidth]{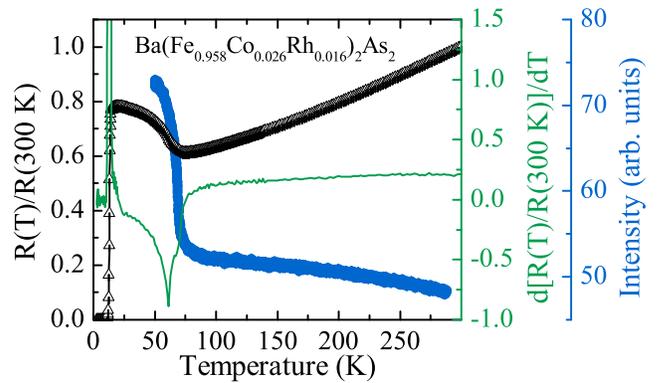} 
\caption{(color online) Normalized resistance (open black triangles), the derivative of the resistance (green line), and the structural order parameter (closed blue circles) for $x + y = 0.042$, Ba(Fe$_{0.958}$Co$_{0.026}$Rh$_{0.016}$)$_2$As$_2$.}
\label{STnRes}
\end{figure}

We present normalized electrical resistance data between $T = 2$ K and 300 K for selected Ba(Fe$_{1-x-y}$Co$_x$Rh$_y$)$_2$As$_2$ compounds in Fig.~\ref{resistivity}. We measured as-grown samples to avoid shaping samples for the resistivity measurement to prevent cracks or exfoliation of the sample\cite{ni08, Canfield09, ni_phase_2009} and normalized our resistance data by the resistance value at $T = 300$ K for each measurement. We find anomalies in the resistance data, which represent $T_\mathrm{S}$ and $T_\mathrm{N}$, as previously seen in transition-metal doped BaFe$_2$As$_2$ compounds\cite{canfield10, Johnston10, Dai15,sefat08, ni08, li_superconductivity_2009, Canfield09, ni_phase_2009, han_2009, saha_superconductivity_2010}. For instance, the resistance anomalies appear at $T = 99.6$ K and $T = 96.5$ K for the sample with $x = 0.026$ and $y = 0.000$, which are consistent with the reported values of $T_\mathrm{S}$ and $T_\mathrm{N}$ for similar compositions, respectively.\cite{ni08,Canfield09} These values are obtained from the derivative of the resistance data and an example of the derivative of the resistance data is shown in Fig.~\ref{STnRes} for Ba(Fe$_{0.958}$Co$_{0.026}$Rh$_{0.016}$)$_2$As$_2$.
These anomalies appear at lower temperatures when more Rh is doped. In the sample with $x = 0.028$ and $y = 0.031$, we no longer see the resistance anomaly which indicates no structural and AFM transitions. The systematic changes of $T_\mathrm{S}$ and $T_\mathrm{N}$ seen from the resistance data for Co-Rh doped Ba(Fe$_{1-x-y}$Co$_x$Rh$_y$)$_2$As$_2$ compounds are consistent with the behaviors in Co122 and Rh122\cite{ni08,Canfield09}. Although it is not yet clear whether Co and Rh donate the same number of extra electrons, we attempt to analyze and understand our data in terms of total electron doping and denote our data using the total doping ($x + y$) in the rest of the paper unless otherwise is necessary. 

\begin{figure}[t!]
\centering
\includegraphics[width=0.95\linewidth]{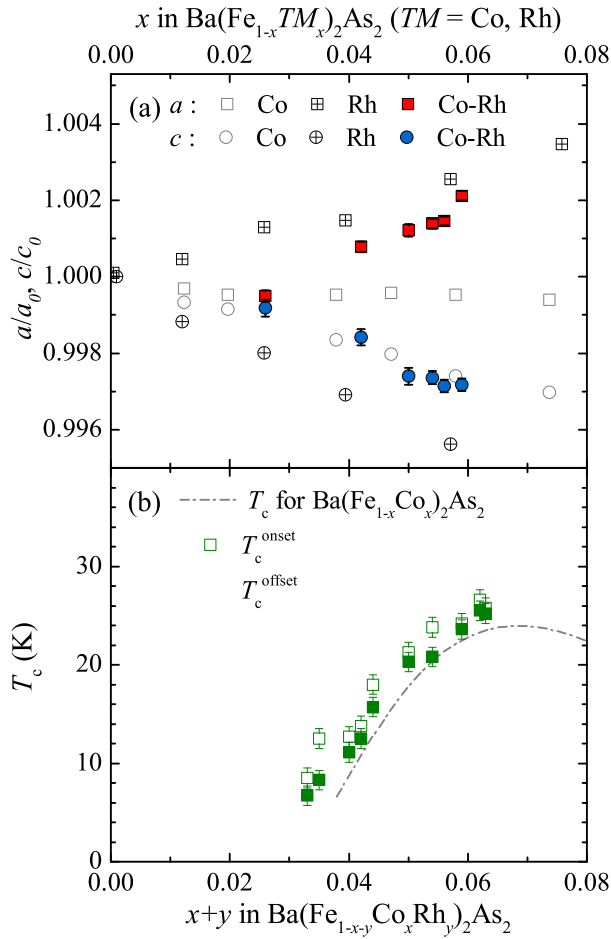} 
\caption{(color online) (a) Normalized lattice parameters. Filled symbols show $a/a_{0}$ (square) and $c/c_{0}$ (circle) for Ba(Fe$_{1-x-y}$Co$_x$Rh$_y$)$_2$As$_2$ as a function of the sum of $x$ and $y$. $a_{0} = 3.9697(1)$ \AA~and $c_{0} = 13.0583(4)$ \AA . Open and crossed symbols are $a/a_{0}$ and $c/c_{0}$ for Ba(Fe$_{1-x}$Co$_x$)$_2$As$_2$ and Ba(Fe$_{1-x}$Rh$_x$)$_2$As$_2$, respectively, from Ref.~\onlinecite{ni_phase_2009}. (b) Superconducting transition temperatures ($T_\mathrm{c}$). Open symbols and closed symbols represent the onset and offset temperatures for Ba(Fe$_{1-x-y}$Co$_x$Rh$_y$)$_2$As$_2$, respectively (see the text for detail). The line indicates $T_\mathrm{c}$ for Ba(Fe$_{1-x}$Co$_x$)$_2$As$_2$ from Ref.~\onlinecite{ni08,Canfield09}.}
\label{lattice}
\end{figure}

Figure~\ref{lattice} (a) shows the lattice parameters $a$ and $c$ at room temperature normalized by the values for the parent BaFe$_2$As$_2$ compound. For the sample with $x = 0.027$ and $y = 0.000$, i.e. Ba(Fe$_{0.973}$Co$_{0.027}$)$_2$As$_2$, the normalized lattice parameters are close to the previously reported values\cite{ni08,Canfield09}. We find that the in-plane lattice parameter $a$ increases whereas the out-of-plane lattice parameter $c$ decreases with increasing Rh doping in Ba(Fe$_{0.973-y}$Co$_{0.027}$Rh$_y$)$_2$As$_2$. We compare our data with those for Co122\cite{ni08,Canfield09} and Rh122\cite{ni_phase_2009}, which are shown with open and crossed symbols, respectively, in Fig.~\ref{lattice} (a). While a slight decrease is observed in the lattice parameter $a$ for Co122, the lattice parameter $a$ for CoRh122 increases significantly and follows the trend in Rh122. In contrast, the lattice parameter $c$ for CoRh122 tracks closely the change in the lattice parameter $c$ in Co122 whereas the lattice parameter $c$ for Rh122 is much larger in all composition ranges. 

Figure~\ref{lattice} (b) presents the superconducting transition temperature ($T_\mathrm{c}$) as a function of electron doping, $x + y$. The onset and offset $T_\mathrm{c}$ were determined from the resistance measurements using the criteria described in Ref.~\onlinecite{Ni_onset_08}. We find that the superconducting transition temperatures for CoRh122 [symbols in Fig.~\ref{lattice} (b)] are very similar to those observed in Co122 and Rh122.\cite{ni08,Canfield09} For comparison, the $T_\mathrm{c}$ phase line for Co122 is shown as the dot-dash line in Fig.~\ref{lattice} (b). Earlier studies have argued that the details of the crystal structure, such as the pnictogen-Fe-pnictogen angle and the pnictogen height, may play a significant role in high-$T_\mathrm{c}$.\cite{Johnston10, Kuroki09, Calderon09} Although we do not precisely know these details, we can deduce that such details are likely different between Co122, Rh122, and CoRh122 based on the behaviors of the lattice parameters $a$ and $c$ in these compounds. Despite this potential difference, superconductivity is surprisingly robust in Co122, Rh122, and CoRh122 as shown in Fig.~\ref{lattice} (b) .

\begin{figure}[t!]
\includegraphics[width=0.95\linewidth]{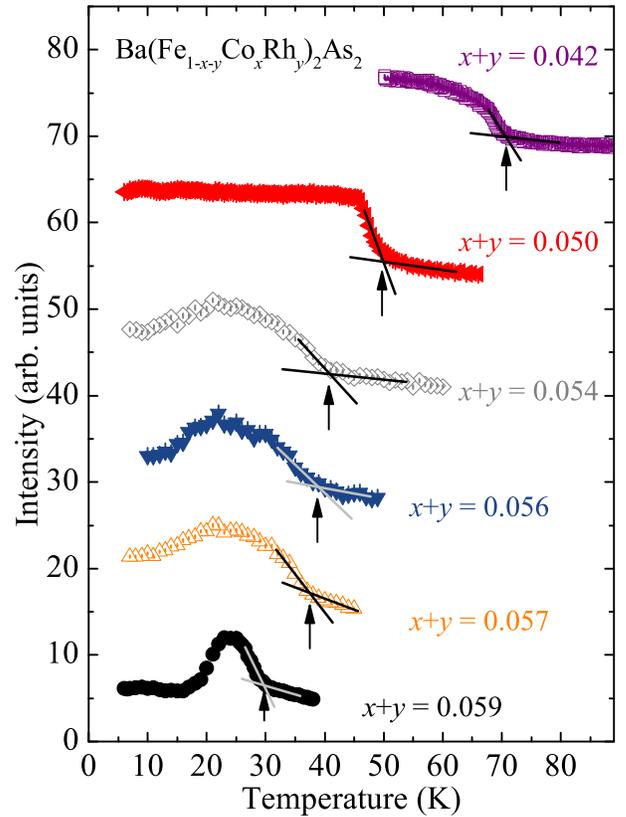} 
\caption{(color online) Changes of the peak intensity at the nuclear (4,~0,~0) Bragg peak as a function of temperature. The structural transition temperature ($T_\mathrm{S}$, the position of arrows) was determined at the point where the peak intensity raises sharply and the resistivity anomaly appears. Note the decrease of the subsequent peak intensity below $T_\mathrm{c}$ in $x + y \geq$ 0.054.}
\label{ST}
\end{figure}

Now we turn to the results of the single-crystal neutron diffraction measurements. We first present the structural order parameters in Fig.~\ref{ST} which were obtained by measuring changes of the peak intensity at the nuclear (4,~0,~0) peak as a function of temperature. The change in the peak intensity is associated with an extinction release across a structural phase transition\cite{Lester09, Kreyssig10,Lu_science_2014}. Measurements of extinction release as a surrogate structural order parameter are very sensitive to the quality of the samples and usually result in various shapes of order parameters (see Figures in Ref.~\onlinecite{Lester09, Kreyssig10,Lu_science_2014,Kim_2015}) which make the determination of $T_\mathrm{S}$ difficult.
So we first determined $T_\mathrm{S}$ from the order parameters at a temperature where the intensity increases sharply. Then we compared this $T_\mathrm{S}$ with the temperature where the resistivity anomaly is observed. An example of this method is shown in the Fig.~\ref{STnRes} for Ba(Fe$_{0.958}$Co$_{0.026}$Rh$_{0.016}$)$_2$As$_2$. Since the values from two different measurements are consistent with each other, we can rely on this method to determine $T_\mathrm{S}$. In Fig.~\ref{ST}, the $T_\mathrm{S}$ is obtained from this method and marked with arrows.

A closer inspection of Fig.~\ref{ST} shows that $T_\mathrm{S}$ decreases continuously with increasing electron (Rh) doping, reaching $T_\mathrm{S} = 30 \pm 1$ K for $x~+~y~=$ 0.059 (Fig.~\ref{ST}), and then disappears at $x~+~y~=~0.061$ (not shown). In samples with $x~+~y~\geq~0.054$, the intensity at the nuclear (4,~0,~0) peak decreases below $T_\mathrm{c}$, which is consistent with the suppression of the orthorhombicity.\cite{nandi_anomalous_2010} As the crystal structure becomes less orthorhombic, a part of the diffracted intensity becomes extinct, resulting in the suppression of the observed intensity. For the $x~+~y~=~0.059$ sample, the peak intensities below $T~\approx~17$ K are almost the same as the value at $T_\mathrm{S}$. This implies that this sample re-enters a tetragonal structure below $T~\approx~17$ K. We conclude that the structure of $x~+~y~=~0.059$ changes first from tetragonal to orthorhombic at $T_\mathrm{S}~=~30$ K and re-enters a tetragonal structure at $T~\approx~17$ K which is below $T_\mathrm{c}$. 

\begin{figure}[t!]
\includegraphics[width=0.95\linewidth]{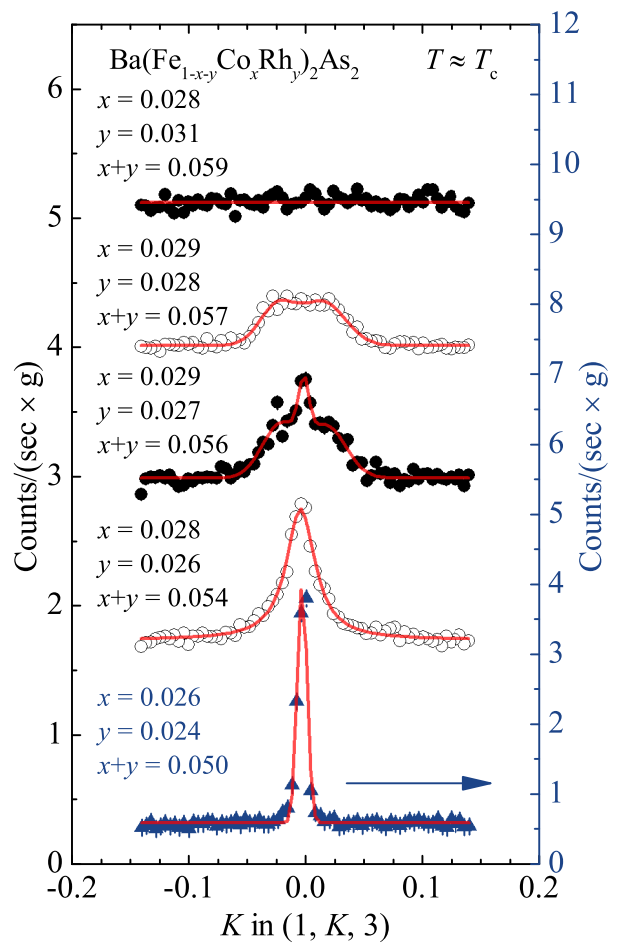}
\caption{Transverse neutron scattering near the (1,~0,~3) magnetic Bragg point at $T \approx T_\mathrm{c}$. Scans are offset vertically and scaled for clarity}
\label{scans}
\end{figure}

Figure~\ref{scans} presents scans along the transverse direction, i.e. the orthorhombic $\bm{b}$ direction, through the (1,~0,~3) AFM Bragg position in Ba(Fe$_{1-x-y}$Co$_x$Rh$_y$)$_2$As$_2$. We plot the scans at $T~\approx~T_\mathrm{c}$ where the AFM signal is maximum. 
A sharp single AFM peak is observed for $x + y \leq 0.050$, which is consistent with the commensurate (C) AFM ordering.
With slightly more electron doping, the peak becomes broad along the orthorhombic $\bm{b}$ direction in $x + y = 0.054$, similar to the observation in other electron doped compounds.\cite{Kim12} Then, three peaks are observed at $x + y = 0.056$. The three peaks consist of one central commensurate (C) peak at $\bm{Q}_\mathrm{AFM}$ and two satellite incommensurate (IC) peaks at $\bm{Q}_\mathrm{AFM} \pm \bm{\tau}$. Observation of three peaks indicates the coexistence of C and IC AFM phases in this sample, which is consistent with the first order C-to-IC transition\cite{Pratt_2011, Kim12}.
With further Rh doping, only IC AFM peaks remain at $T \approx T_\mathrm{c}$ for $x + y = 0.057$.
Finally, we no longer detect any signals around $\bm{Q}_\mathrm{AFM}$ for $x~+~y~=~0.059$ and conclude that the AFM ordering is completely suppressed in samples with $x~+~y~\geq~0.059$.
The smooth evolution of the AFM ordering and the first order C-to-IC transition are consistent with the behavior seen in superconducting Co or Ni doped compounds.\cite{Pratt_2011, Kim12} In addition, the critical concentration, $x + y =$ 0.056, of a first-order C-to-IC transition in Ba(Fe$_{1-x-y}$Co$_{x}$Rh$_{y}$)$_2$As$_2$ is the same as the value ($x_\mathrm{c} = 0.056$) observed for Co122.\cite{Pratt_2011}
We fit the scans with a single Gaussian peak for $x + y = 0.050$, a single Lorentzian for $x + y = 0.054$, three Gaussian peaks for $x + y = 0.056$, and two Gaussian peaks for $x + y = 0.057$ and show the results of the best fits with lines in Fig.~\ref{scans}. From the fits for $x + y = 0.056$ and 0.057, we find that the incommensurability, $\tau$, for both compounds is $0.020 \pm 0.002$ reciprocal lattice units (r.l.u.) which are identical within the error. This value is slightly smaller than the values for the single Co ($\tau \approx 0.025 - 0.030$) or Ni ($\tau = 0.033$) doped compounds\cite{Pratt_2011, Kim12}. 

\begin{figure}[t!]
\includegraphics[width=0.95\linewidth]{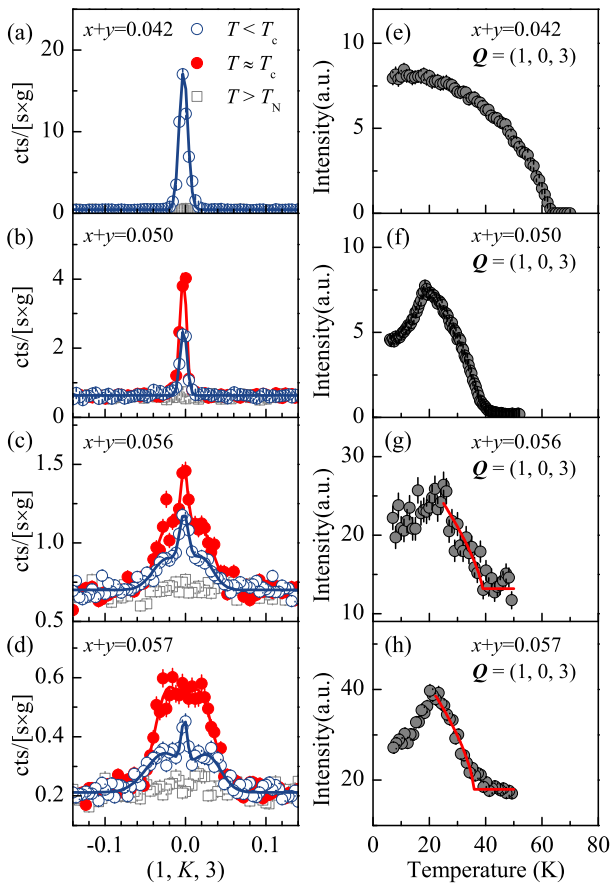} 
\caption{(color online) Transverse neutron diffraction scans through the (1,~0,~3) magnetic Bragg peak at temperature $T < T_\mathrm{c}$ (open blue circles), $T \approx T_\mathrm{c}$ (closed red circles), and $T > T_\mathrm{N}$ (open gray rectangles) for Ba(Fe$_{1-x-y}$Co$_x$Rh$_y$)$_2$As$_2$ with (a) $x + y =$ 0.042, (b) 0.050, (c) 0.056, and (d) 0.057. The corresponding AFM order parameters are shown in (e)-(h). Lines are guides to eyes.}
\label{AFM}
\end{figure}

In order to study the temperature dependence of the AFM ordering, we plot transverse scans at three different temperature regimes and the corresponding order parameters in Fig.~\ref{AFM}. 
For $x + y = 0.042$ and 0.050, a single sharp AFM peak exists down to the lowest temperature [Fig.~\ref{AFM} (a) and (b)]. While the AFM peak for $x + y = 0.042$ increases continuously [Fig.~\ref{AFM} (e)], the intensity of the peak for $x + y = 0.050$ increases first then decreases below $T_\mathrm{c}$ [Fig.~\ref{AFM} (f)]. 
For $x + y = 0.056$, three AFM peaks are observed at all temperatures below $T_\mathrm{N}$ [Fig.~\ref{AFM} (c)]. As temperature is lowered through $T_\mathrm{c}$, the order parameter measured at the C AFM position is suppressed [Fig.~\ref{AFM} (g)]. When we compare the scans between $T \approx T_\mathrm{c}$ [closed red circles in Fig.~\ref{AFM} (c)] and $T < T_\mathrm{c}$ [open blue circles in Fig.~\ref{AFM} (c)], we find that the intensities of the C and IC peaks decrease at a similar rate below $T_\mathrm{c}$. 
For the compound with $x + y =0.057$, we observed only two IC AFM peaks at $T \approx T_\mathrm{c}$ (Fig.~\ref{scans}). At $T < T_\mathrm{c}$, the order parameter measured at the C AFM position is suppressed as expected from the competition between magnetism and superconductivity [Fig.~\ref{AFM} (h)]. However, we observe three AFM peaks at $T < T_\mathrm{c}$ in the compound with $x + y =$ 0.057. It is likely that the central C AFM peak is present but not distinguishable at $T \approx T_\mathrm{c}$[Fig.~\ref{AFM} (d)].
By looking at the intensity changes across $T_\mathrm{c}$ between the central C peak and the satellite IC peaks, we find that the suppression of the IC peaks is greater than that of the C AFM peak. 
This observation suggests that the C AFM may be more stable than the IC AFM in the competition with superconductivity. It is interesting to note that the non-superconducting Cu doped Ba(Fe$_{1-x}$Cu$_x$)$_2$As$_2$ compounds exhibit commensurate AFM ordering in the entire composition range. Since we only have detailed $Q$ scans at three temperatures, as presented here, and the AFM order parameters were measured at the $\bm{Q}_\mathrm{AFM}$ position, further studies are required to understand this behavior.

\begin{figure}[t!]
\includegraphics[width=0.95\linewidth]{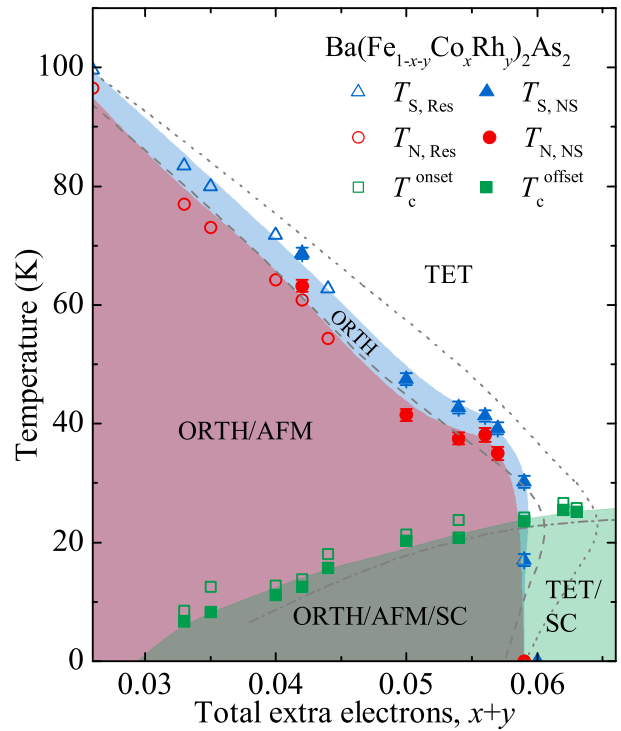} 
\caption{ (color online) Experimental phase diagram for Ba(Fe$_{1-x-y}$Co$_x$Rh$_y$)$_2$As$_2$ determined from neutron diffraction (closed triangles and circles) and transport measurements (open triangles, circles, rectangles, and closed rectangles) as well as the data from the single Co doping (gray lines)\cite{ni08,Canfield09,Fernandes10,nandi_anomalous_2010}. Tetragonal (Tet), orthorhombic (Orth), antiferromangetic (AFM), and superconducting (SC) phases are noted and color-coded. The re-entrance temperature from the orthorhombic to tetragonal phase for $x + y = 0.059$ is denoted with a half-filled triangle. }
\label{PD}
\end{figure}

\section{Summary}

We summarize our results in the phase diagram of Ba(Fe$_{1-x-y}$Co$_x$Rh$_y$)$_2$As$_2$ compounds in Fig.~\ref{PD}. The phase diagram is constructed from the transport and neutron measurements together with the phase lines of Co122. We see that while the AFM phase transition temperatures in CoRh122 are comparable to the values for Co122, the structural transition temperatures are lower for CoRh122. Consequently, the difference between $T_\mathrm{S}$ and $T_\mathrm{N}$ is smaller for CoRh122. At higher doping level, both the structural and AFM phase transitions terminate at about $x + y = 0.059$ in Ba(Fe$_{1-x-y}$Co$_x$Rh$_y$)$_2$As$_2$, which is smaller than $x \approx 0.06$ and $0.064$ (for AFM and structural phase transitions, respectively) for Ba(Fe$_{1-x}$Co$_x$)$_2$As$_2$. However, the critical concentration for the C-to-IC AFM transition is similar in CoRh122 and Co122. The back-bending of both the structural and AFM phase lines, observed in Co122 (dotted and dashed lines, respectively, in Fig.~\ref{PD}), are not clearly present in CoRh122. In addition the back-bending was not observed for the Ni doped compounds. Instead, the phase lines for the Ni doped compounds disappeared very suddenly, resulting in an avoided quantum critical point\cite{Lu13}, which may be the case for CoRh122. We observed surprising agreement between the superconducting transition temperatures in CoRh122 and Co122 while the details of structure, seen from systematic measurements of the lattice parameters, are different in the two compounds. This indicates that the electron doping plays the essential role in determining the $T_\mathrm{c}$ in this family of FeAs-based compounds.

Taken together, we have shown that the changes in the structural and antiferromagnetic phase transitions, suppression of their order parameters below $T_\mathrm{c}$, and the emergence of superconductivity in Ba(Fe$_{1-x-y}$Co$_x$Rh$_y$)$_2$As$_2$ compounds are very similar to those in Ba(Fe$_{1-x}$Co$_x$)$_2$As$_2$ compounds whereas the fine details of those properties are slightly different between both materials. This clearly indicates that a simple rigid band picture works well in explaining the overall properties of electron doped superconducting BaFe$_2$As$_2$ compounds including Ba(Fe$_{1-x-y}$Co$_x$Rh$_y$)$_2$As$_2$ compounds.

\begin{acknowledgments} 

We are grateful to M. Wang and Z. Xu for valuable discussions. The work at the Lawrence Berkeley National Laboratory was supported by the U.S. Department of Energy (DOE), Office of Basic Energy Sciences, Materials Sciences and Engineering Division, under Contract No. DE-AC02-05CH11231. 
\end{acknowledgments}

\bibliographystyle{apsrev4-1}
\bibliography{Rh_ins}

\end{document}